# Microlensing by Stars

Marc Kamionkowski[1]
School of Natural Sciences, Institute for Advanced Study, Princeton, NJ 08540
Physics Department, Columbia University, New York, New York 10027

## ABSTRACT

If stars at the lower end of the main sequence are responsible for the microlensing events observed in the Galactic bulge, then light from the lensing star contributes to the observed brightness. The background and lensing stars generally have different colors, and the relative brightness changes during the microlensing event. Therefore, microlensing light curves are not perfectly achromatic if hydrogen-burning stars are the lenses. In most cases, the color shift will be too small to be observable, but we argue that given the current microlensing rates, it is plausible that a few color-shifted microlensing events could be observed in the near future, especially if strategies are optimized to search for them. Although rare, such an event could potentially provide a wealth of information: Light curves in two bands can be used to determine the masses and distances of the two stars as well as the transverse speed of the lensing star. Light curves in additional wavebands could make the determination more precise.

*Subject headings:* Galaxy: general – Galaxy: structure — gravitational lensing

## 1. INTRODUCTION

Microlensing experiments were first proposed as a means of probing the Galactic halo for massive compact halo objects (MACHOs), non-luminous objects with masses within a handful of orders of magnitude of the solar mass, $\mathcal{M}_\odot$, for example, black holes, neutron stars, or stars with masses below the hydrogen-burning threshold (Paczyński 1986; Griest 1991). If the halo is composed of MACHOs, then observation of several million stars in the LMC would produce a significant number of microlensing events. Although such events have been observed (Aubourg et al. 1993; Alcock et al. 1994), the rates appear to be too low to account for the dark matter needed to support rotational velocities at large radii. It was similarly proposed that observation of the Galactic center could be used to study disk dark matter and to probe the mass function of stars too faint to be observed optically.

---

[1]kamion@phys.columbia.edu

Both the OGLE (Udalski et al 1994a) and MACHO (Alcock et al. 1994) collaborations have recently reported observation of numerous candidate microlensing events toward the Galactic center. The OGLE estimate of the optical depth to microlensing of Galactic bulge stars in Baade's window is $\sim (3.3 \pm 1.2) \times 10^{-6}$, and the MACHO results are consistent. This is significantly higher than the optical depth expected from microlensing by faint stars in the disk or bulge, or from MACHOs which account for the disk dark matter; these theoretical estimates generally fall in the range $(0.5 - 1.0) \times 10^{-6}$ (Paczyński 1991; Griest et al. 1991; Kiraga & Paczyński 1994; Giudice, Mollerach & Roulet 1994; Han & Gould 1994).

Kiraga & Paczyński (1994) suggested that the enhancement could be due to a bar in the Galactic center pointed at us. Using a self-consistent model for the bar (Zhao 1994) which matches kinematic observations of the bulge (Zhao, Spergel & Rich 1994a) as well as the COBE image of the Galaxy, the optical depth to microlensing was calculated (Zhao, Spergel & Rich 1994b). It was found that the optical depth from lensing by dwarfs with a mass function $dn/d\mathcal{M} \propto \mathcal{M}^{-2}$ in this model was consistent with the OGLE and MACHO results. Furthermore, the best fit to the OGLE time-duration distribution was obtained with a lower-mass cutoff of the lens mass function of $0.1 \mathcal{M}_\odot$, and the mean lens mass was found to be $0.4 \mathcal{M}_\odot$, i.e., *above* the hydrogen-burning threshold. Most of the lenses are found to be in the bulge with a line-of-sight distance 6.25 kpc, distances comparable to, but slightly smaller, than the stars being lensed.

If the lensing objects are ordinary stars at the lower end of the main sequence, then the observed light comes from the lens as well as from the background star being amplified. The two stars will generally have different colors, so the microlensing light curves should not be perfectly achromatic. In fact, since the majority of lenses are much fainter than the background stars being monitored, most of the light curves are effectively achromatic. However, in some fraction of the events, the brightness of the lens should be comparable to or greater than the brightness of the background star being lensed, in which case there could be an observable color shift. In this Letter, I argue that measurement of light curves in two (or more) wavebands in a color-shifted microlensing event can be used to determine the masses and distances to the background and lensing star, as well as the transverse speed of the lens through the microlensing tube. Thus observation of a few of these events could potentially provide as much information on the lensing objects as numerous achromatic events.

In the following Section, I describe color-shifted microlensing events. In Section 3, I estimate the rates for color-shifted events. Although small, the fraction of events which are detectably color-shifted may be large enough that such events could be observed in the near future, especially given the unexpectedly large rate of achromatic events observed so far. In Section 4, I explain how masses, distances, and transverse velocities can be obtained from color-shifted light curves in two (or more) bands, and some concluding remarks are made in the final Section.

## 2. COLOR-SHIFTED MICROLENSING EVENTS

Assume that the lensing and lensed stars are both on the main sequence. Giants have large radii and are generally brighter than dwarfs, so it unlikely that a giant will be microlensed, and even less likely that a giant will act as a lens. Then, the luminosity of a star of mass $\mathcal{M}$ is roughly $(L/L_\odot) \simeq (\mathcal{M}/\mathcal{M}_\odot)^3$, and the brightness of a star at a distance $d$ is $\ell = L/d^2$. Furthermore, the color of a star (found, e.g., by measuring the brightness in two different wavebands) determines the star's spectral class and therefore its mass and luminosity.

First, let us review standard microlensing of a star by a nonluminous (or faint) lensing star. In most microlensing events, a background star with mass $\mathcal{M}_b$ at a distance $d_b$ is lensed by a foreground star of mass $\mathcal{M}_f \ll \mathcal{M}_b$ and at a distance $d_f < d_b$ such that the brightness of the lens is smaller than that of the background star, $\ell_f \lesssim \ell_b$. So, the foreground star is too faint to be distinguished. If the distance of the lensing star from the lensed-star line of sight is $R_e u$, where (Griest 1991)

$$R_e = \left[\frac{4G\mathcal{M}_f d_f (d_b - d_f)}{c^2 d_b}\right]^{1/2} = 3.2 \text{ a.u.} \left[\frac{\mathcal{M}_f}{\mathcal{M}_\odot} \frac{d_b}{8 \text{ kpc}} \frac{x'}{0.8} \frac{1-x'}{0.2}\right]^{1/2} \quad (1)$$

is the Einstein radius of the lens and $x' = d_f/d_b$, then the amplification of the background star as a function of time $t$ is

$$A[u(t)] = \frac{u^2 + 2}{u(u^2 + 4)^{1/2}}, \qquad u(t) = [\omega^2 (t - t_0)^2 + u_{\min}^2]^{1/2}, \quad (2)$$

where $u_{\min}$ is the impact parameter in units of the Einstein radius, $\omega = v_\perp R_e^{-1}$ and $v_\perp$ is the transverse speed of the lens through the microlensing tube, and $t_0$ is the time at which peak amplification, $A_{\max} = A(u_{\min})$, occurs.

A microlensing event is registered when the amplification of an event exceeds a threshold $A_T$ which corresponds to a dimensionless lens–line-of-sight distance of $u_T$ (e.g., $A_T = 1.35$ for $u_T = 1$). An achromatic event is described by the three parameters, $u_{\min}$ (or equivalently, $A_{\max}$), the timescale $\omega^{-1}$, and the time $t_0$. The event duration $t_e$ is the time that the amplification is above threshold, and it is also the time the lens remains within a distance $u_T R_e$ from the line of sight; it is related to the timescale $\omega^{-1}$. The system parameters (the distances and masses of both stars and the transverse speed) cannot be determined uniquely in any given event. These quantities can in principle be disentangled in a statistical manner if a number of events are observed, and only with some assumptions about the mass, spatial, and speed distributions of the lenses. Microlensing is a gravitational effect, so the amplification is the same in all wavelengths. Therefore, if the lens is too faint to be observed, the microlensing event will be achromatic.

Now consider the more general case where the brightness of the lens is comparable to or greater than that of the background star, $l_f \gtrsim l_b$. If the lenses responsible for the observed events are stars at the lower end of the main sequence, then there should be similar (although rarer) events where more massive main-sequence stars act as lenses. The light observed is a combination of the light from both stars, and the colors of the two stars will generally be different.[2] When the foreground star passes within the lensing tube, the background star will be amplified, and the relative brightnesses will change. Consequently, the color will change, and achromaticity is lost. To properly describe the event, we must consider the brightness in each waveband separately. If the amplification of the background star is $A$, then the brightness in waveband $\lambda$ is

$$\ell_\lambda = \ell_{f\lambda} + A\ell_{b\lambda}, \tag{3}$$

where $\ell_{f\lambda} = L_{f\lambda}/d_f^2$ is the brightness of the foreground star in $\lambda$, $L_{f\lambda}$ is the luminosity of the star in $\lambda$, and similarly for the background star. The baseline brightnesses are obtained by setting $A = 1$ in Eq. (3).

Although $A$ is the amplification of the brightness of the background star, the observed light in this case comes from both stars, so the *observed* amplification, which I will denote by $\mathcal{A}$, is different from the microlensing amplification $A$. Although the amplification of the background star is indeed achromatic, the observed amplification will depend on

---

[2] The analysis here is similar to that for lensing of binary stars by MACHOs as discussed by Griest & Hu (1992). Light from an additional unresolved star was also considered by Gould & Loeb (1992) and Gould (1994), and was included in the analysis of the possible binary microlens detected by the OGLE collaboration (Udalski et al. 1994b).

wavelength, and in waveband $\lambda$ it is

$$\mathcal{A}_\lambda(t) = \frac{\ell_{f\lambda} + A(t)\ell_{b\lambda}}{\ell_{f\lambda} + \ell_{b\lambda}} = (1 - r_\lambda) + A(t)r_\lambda, \qquad (4)$$

where $r_\lambda = \ell_{b\lambda}/(\ell_{b\lambda} + \ell_{f\lambda})$ (Griest & Hu 1992). Suppose, for example, light curves are measured in two bands, $\alpha$ and $\beta$. If $\ell_f \gtrsim \ell_b$ (in both bands), then the baseline color observed will be $\ell_{f\alpha}/\ell_{f\beta}$ (the color of the lens), but in a lensing event with amplification $A \gg 1$, the observed color will be $\ell_{b\alpha}/\ell_{b\beta}$ (the color of the background star).

Such a color-shifted microlensing event will be registered when the *observed* amplification is greater than threshold, $\mathcal{A} > A_T$. The observed amplification depends on the band, so strictly speaking, the exact time at which an event is registered (or whether it is registered at all) may depend on the waveband. If events are triggered by the light curve in $\lambda$, they will be registered when the microlensing amplification is

$$A \geq A_{\text{thresh}} = \frac{A_T - (1 - r_\lambda)}{r_\lambda}, \qquad (5)$$

or only when the dimensionless lens–line-of-sight distance is $u \leq u_{\text{thresh}} = u(A_{\text{thresh}})$.

The duration of a color-shifted event is $t_e = 2[u_{\text{thresh}}^2 - u_{\text{min}}^2]^{1/2}\omega^{-1}$. The duration of any given event depends on the transverse speed, the mass of the lens, and the distances to both stars through $\omega$, and it depends on the impact parameter and the luminosities and distances to both stars through the $r$ dependence of $u_{\text{thresh}}$. In general, however, the duration of the color-shifted events should be shorter than the duration of achromatic events. First note that $R_e \propto \mathcal{M}_f^{1/2}$, and that $u_{\text{thresh}} \propto (\ell_b/\ell_f) \propto (\mathcal{M}_b/\mathcal{M}_f)^3$ for $\mathcal{M}_b \ll \mathcal{M}_f$ (and a mass-luminosity relation $L \propto \mathcal{M}^3$). Fix the distances of the two stars, the impact parameter, and the transverse speed. Then consider the duration of an achromatic event ($\ell_f \ll \ell_b$ and $\mathcal{M}_f \ll \mathcal{M}_b$) relative to the duration of an event where the lens mass is larger than that of the background star ($\ell_f \gg \ell_b$). Although the Einstein radius is larger in the latter case (which would make the duration longer), the fraction of the Einstein radius that gives rise to an observed amplification above threshold is smaller, and the latter effect dominates.

## 3. RATE OF COLOR-SHIFTED EVENTS

It is easy to see that color-shifted events will be rare compared to achromatic events for two reasons. First, color-shifted events require lenses with masses which are generally larger than those in achromatic events, and the mass function decreases with increasing mass, at least by assumption in the Zhao et al. (1994a,b) model. Second, although the Einstein radius is larger for a larger-mass lens, the cross section for a color-shifted event is generally smaller due to the fact that the fraction of the Einstein radius that gives rise to an observed amplification greater than threshold is reduced. Even so, it is still plausible that a handful of color-shifted events can be observed in the near future given the large achromatic-event rates reported so far (Udalski et al. 1994a; Alcock et al. 1994) and the estimate for the fraction of color-shifted events provided in this Section.

To estimate the rate of color-shifted events, assume that the observed microlensing events are all due to lensing by faint main-sequence stars in the bulge. Strictly speaking, all events are then color shifted. However, if the mass of the background star is greater than the mass of the lens, $\mathcal{M}_b > \mathcal{M}_f$, then the lensing event will most likely appear to be achromatic in current experiments. On the other hand, if the lens mass is larger, $\mathcal{M}_f > \mathcal{M}_b$, then the observed light at baseline comes primarily from the lensing star and the light at peak amplification comes primarily from the background star, so it is plausible that there will be an observable color shift. If the lensing stars are in the bulge, then the distances to the lenses are comparable to the distances to the background star (Zhao, Spergel & Rich 1994b). Take the mass function to be $dn/d\mathcal{M} \propto \mathcal{M}^{-2}$.

The rate for achromatic events is

$$R_{\text{achrom}} \propto \int_{\mathcal{M}_{\text{min}}} d\mathcal{M}_b \int_{\mathcal{M}_{\text{cut}}}^{\mathcal{M}_b} d\mathcal{M}_f \frac{dn}{d\mathcal{M}_b} \frac{dn}{d\mathcal{M}_f} R_e^2(\mathcal{M}_f), \tag{6}$$

while the rate for color-shifted events is

$$R_{\text{cs}} \propto \int_{\mathcal{M}_{\text{min}}} d\mathcal{M}_f \int_{0.1\mathcal{M}_\odot}^{\mathcal{M}_f} d\mathcal{M}_b \frac{dn}{d\mathcal{M}_b} \frac{dn}{d\mathcal{M}_f} R_e^2(\mathcal{M}_f) u_{\text{thresh}}^2(\mathcal{M}_b, \mathcal{M}_f), \tag{7}$$

where $\mathcal{M}_{\text{min}}$ is the mass of the faintest star (in the bulge) observable in the experiment, $\mathcal{M}_{\text{cut}}$ is the smallest stellar mass (i.e., where the mass function drops to zero), and $R_e(\mathcal{M}_f) \propto \mathcal{M}_f^{1/2}$. The upper limit to the first integral in Eq. (6) is of no consequence. The lower limit, $0.1\mathcal{M}_\odot$, in Eq. (7) is the smallest mass at which a star undergoes hydrogen burning and is therefore luminous. The function $u_{\text{thresh}}(\mathcal{M}_b, \mathcal{M}_f)$ can be obtained from the formulas in the previous section and taking $L \propto \mathcal{M}^3$ and $d_f \simeq d_b$. The maximum amplification due to the finite size of the source is roughly $2R_e/r_b$, where $r_b$ is the radius of the lensed star, and this leads to a lower limit to $u_{\text{thresh}}$.[3] The ratio of color-shifted events to achromatic events is then estimated to be roughly

$$\frac{R_{\text{cs}}}{R_{\text{achrom}}} \sim \frac{0.4}{\ln(\mathcal{M}_{\text{min}}/\mathcal{M}_{\text{cut}})}. \tag{8}$$

---

[3]I thank D. Bennett for pointing this out.

If the faintest stars monitored in current experiments are near the top of the main sequence, with masses $\mathcal{O}(10\mathcal{M}_\odot)$, say, and we take the smallest stellar mass to be the cutoff $(0.1\mathcal{M}_\odot)$ suggested by Zhao, Spergel & Rich (1994b), then there should be roughly one color-shifted event for every $\mathcal{O}(10)$ achromatic events.

The fraction depends logarithmically on the mass of the faintest observable star, $\mathcal{M}_{\min}$ and increases as this mass is decreased, as it should. Unlike the case for microlensing by MACHOs, the probability for color-shifted microlensing is *not* independent of the star's spectral class. The probability for a color-shifted event increases if fainter objects are monitored, as the number of objects that could give rise to a signal above threshold increases. Therefore an observing program that goes deeper in magnitude could potentially increase the fraction of color-shifted events as well as the total number of lenses observed.

Of course, to provide a more precise estimate of the fraction of events in which there is an observable color shift, the bands in which light curves are measured as well as a minimum color shift must be specified, and a color-magnitude relation for stars on the main sequence must be employed. Although I have not yet done such a calculation carefully, we can adapt the results of the calculation for a related type of event. Griest & Hu (1992) considered microlensing of binary sources by MACHOs. As in the case considered here, lensing of a background star by an ordinary star, the observed light from a binary source is a combination of the light from the two sources, and if the fainter object is lensed, there may be an observable color shift. If the mass function of stars in binaries is the same as that for single stars, then the fraction of events which are color shifted that they find should be similar to the fraction of color-shifted events which occur when ordinary stars lens ordinary stars. They estimated the fraction of events in which there would be a color shift greater than 0.1 mag to be about $2\% - 5\%$, a rate consistent in order of magnitude with the estimate above. As they point out, the fraction is small because it is rare to find two stars in a binary with comparable brightness and different colors. If lensing occurs by dwarf stars, then the distances will generally be different, and the chance of finding two stars with comparable brightness but different colors is greater. Therefore, the fraction of events where stars are the lens should be slightly higher than the estimate of Griest & Hu (1992).

## 4. MASSES, DISTANCES, AND TRANSVERSE VELOCITIES

So far, we have seen that if faint stars are responsible for the observed microlensing events, then there should be events in which the lens is not too faint which give rise to color-shifted light curves. The rate for such events should be small, and it may require some effort to actually observe these. One might also wonder whether color shifts that arise from lensing by ordinary stars can be distinguished from lensing of binary sources by MACHOs. Although rare and perhaps elusive, observation of a color-shifted microlensing event would be more than just a curiosity: Sufficiently accurate measurement of color-shifted microlensing light curves in two (or more) bands can potentially break the degeneracy between the undetermined system parameters that occurs in achromatic microlensing events: In particular, the masses and distances of the two stars, and the transverse speed of the lensing object (relative to the microlensing tube) can be determined, at least in principle. In addition, such events could be distinguished from lensing of binary sources by MACHOs.

Consider first how the mass and distance of an ordinary main-sequence star is determined. If the brightness is measured in two bands, $\alpha$ and $\beta$, say, then the color, $\ell_\alpha/\ell_\beta$, determines the spectral class, and therefore the mass and luminosity (in any band) of the star. The distance is then obtained by comparing the luminosity with the measured brightness in either (or both) bands. In other words, the two unknowns (mass and distance) are determined by the two observations (brightness in $\alpha$ and $\beta$). Measurements in other bands as well as additional spectral information can be used to break the possible degeneracy between dwarfs and giants, and to account for the effects of interstellar absorption, etc.

If color-shifted microlensing events are observed, similar arguments can give us the masses and distances to the lensing and background stars. An achromatic light curve is fit by three parameters: $\omega$, $t_0$, and $A_{\max}$. On the other hand, the more general color-shifted light curve is determined by four parameters, the additional parameter being the brightness ratio $r_\lambda$. If measured precisely enough, a fit to the light curve yields $r_\lambda$, which together with the observed baseline brightness, $\ell_\lambda$, determines the brightnesses $l_{f\lambda}$ and $l_{b\lambda}$ of both stars. The masses and distances to both stars are then determined from the brightnesses in two bands. Moreover, by demanding that the lensing amplification $A(t)$ be the same in both bands, the statistical significance of the fit can be improved. One need not worry about the possible degeneracy between giants and dwarfs since it can safely be assumed that both stars are on the main sequence and not giants. The relatively large radii of giants makes it unlikely that the lensed star is a giant. It is even less likely that a dwarf would be lensed by a giant.

Once the mass of the lens and distances to both stars are determined, the Einstein radius is known. The transverse speed of the lensing star through the microlensing tube can then be determined from the Einstein radius and the value of $\omega$ obtained from the fits to the light curves. Furthermore, if it is found that $d_f \neq d_b$, then the event cannot be lensing of a binary source by a MACHO.

Additional spectral information can also potentially make the determination more precise. In fact, measurement of the baseline brightness alone in four or more bands provides, in principle, enough information to determine the masses and distances to both stars. In addition, shifts in the relative strengths of spectral lines characteristic of the two stars could be similarly used (Spergel 1994).

## 5. DISCUSSION

It is quite plausible (if not likely) that the microlensing events observed towards the Galactic bulge can be explained by low-mass main-sequence stars in the bulge, and to a lesser extent, in the disk. Although estimates of the event rate expected assuming axisymmetry are smaller than the observed rate, more precise calculations which take into account the deviations from axisymmetry that likely exist in the bulge find rates consistent with observation (Zhao, Spergel & Rich 1994b). In the vast majority of events, the lensing star is much fainter than the background star, so the microlensing event is effectively achromatic. On the other hand, in a small fraction of the events, the brightness of the lensing star may be comparable to or greater than the brightness of the background star being lensed, in which case there should be an observable color shift during the event. Although rare, it is quite plausible that such events could be observed in the near future given the unexpectedly large rate of microlensing events observed so far. Color-shifted events could provide valuable information: Measurements of the light curves in two (or more) bands can be used to determine the masses and distances to both stars, as well as the transverse speed of the lensing star.

It is still not clear whether such color-shifted events are observable by MACHO, the only current experiment monitoring the bulge with light curves in two bands. It is likely that the two bands, blue and red, monitored by MACHO are not ideal for detecting color-shifted events. Color shifts may be more pronounced in other bands (Kamionkowski, Rich, Spergel & Zhao 1994). If so, the current experiments may need to be augmented to increase the sensitivity to color-shifted events. Another possibility is that quick responses by other telescopes to "early warning" systems which notify of a microlensing event in progress could provide the required spectral information during the event.[4] Measurement of the strength of spectral lines during the microlensing event could also provide data which could make the mass and distance determinations more precise (Spergel 1994).

---

[4]There have already been measurements of spectral lines by other telescopes after detection of a microlensing event on the rise (Alcock & Griest, private communication, 1994).

– 10 –

There may be additional background events, e.g., from variable stars, that may mimic color-shifted microlensing events. If so, microlensing events can be distinguished by the symmetry of the light curves about $t_0$ and by demanding that the fits for $A(t)$ obtained in both (or all) bands agree with the standard microlensing light curves, Eq. (2), and with each other.

The exact rate of events will depend not only on a detailed model of the bulge, but also on the catalog of stars monitored by any given experiment. Some or all of the lenses may be low-mass disk stars. If so, the fraction of events that are color shifted should be greater than that if all the lenses are in the bulge. Calculation of the time-duration distribution of color-shifted microlensing events is needed to evaluate the event rate in any given experiment. Such a calculation can be performed reliably only with a detailed model of the bulge and disk that reproduces the microlensing events observed so far (Kamionkowski et al. 1994).

Finally, it should be possible to place an upper limit to the color shifts in the MACHO events observed so far. These could be used to provide an upper limit to the brightness (and with some assumptions about the distance, the mass) of the lens, and thus to the transverse speed, although these limits are most likely relatively weak with current observations.

I thank C. Alcock, J. Bahcall, D. Bennett, A. Gould, K. Griest, A. Loeb, M. Rich, P. Sackett, D. Spergel, and H. Zhao for discussions and comments on a preliminary draft. This work was supported at the I.A.S by the W. M. Keck Foundation, and at Columbia University by the U. S. Department of Energy under contract DE-FG02-92ER40699.